\documentclass[numreferences]{kluwer}

\begin{document}                                                                                   
\begin{article}
\begin{opening}         
\title{Possible Evidence of Time Variation of Weak Interaction 
Constant from Double Beta Decay Experiments}
\author{A.S. \surname{Barabash} \email{barabash@vxitep.itep.ru}}
\institute{Institute of Theoretical and Experimental Physics, B.\
Cheremushkinskaya 25, 117259 Moscow, Russia}
\date{ }
\begin{abstract}
A comparison is made of the probability of the process of
two neutrino double beta decay for $^{82}$Se in direct 
(counter) and geochemical experiments. It is shown that the probability is 
systematically lower in geochemical experiments, which 
characterize the probability of $\beta\beta(2\nu)$ decay a few 
billions          
years ago. The experimental data for 
$^{130}$Te are also analyzed. It is shown that geochemical measurements on young minerals give lower 
values of $T_{1/2} $($^{130}$Te) and $T_{1/2} $($^{82}$Se)
as compared to measurements on 
old minerals. It is proposed that this could be due to a change in 
the weak interaction constant with time. Possibilities 
of new, precise measurements be performed with the aid of 
counters and geochemical experiments are discussed.  

\end{abstract}

\end{opening}

\section{Introduction}
     The question of the dependence of the fundamental constants 
on time was formulated by P. Dirac in 1937 - this is so-called 
Large Number Hypothesis \cite{dira37}. Although Dirac's hypothesis was not 
confirmed in its initial form, interest in this problem gathered 
new strength in the 1980s-1990s, since a time dependence of the coupling 
constants appears in multidimensional Kaluza-Klein-like models 
\cite{chod80,marc84} and in superstring theories 
\cite{wu84,kolb86,grie89,damo94}. 
Recently scheme with time variation of the velocity of light in 
vacuum, c, and the Newtonian gravitation constant $G_N$ was 
proposed 
as a solution of cosmological puzzles  and as possible alternative 
to inflationary cosmology \cite{albr99,barr99,clay99,avel99}.

 On the other hand, it was found some indication that the 
fine structure constant $\alpha$ was smaller at earlier epochs, $
\Delta\alpha/\alpha = (-0.72\pm 0.18)\cdot 10^{-5}$ for redshifts 
$0.5 < z < 3.5$ \cite{webb01}. But 
further work is required to explore possible systematic errors in 
the data.

      So, one can conclude that there are theoretical and 
experimental motivations to search for time variations in the 
fundamental constants.

In this report I present the situation in double beta decay 
where dependence of double beta decay rate with time was recently 
indicated \cite{bara98,bara00}.

\section{Present limit on weak interaction constant time 
variation}

     Modern limits on the possible variations of 
different fundamental constants with time can be found in review 
\cite{uzan02}. 
For example, the strictest limits for the 
weak interaction constant were obtained from an analysis of the 
operation of the natural nuclear reactor in Oklo\footnote{The 
first analysis of the Oklo data for a possible change in the 
fundamental constants with time was done in Ref. \cite{shly76}}: 
$\mid\Delta G_F\mid/G_F < 0.02$ (where  $\Delta G_F  = 
G_F^{\rm Oclo} - G_F^{\rm now}$) or  $\mid\dot G_F/G_F\mid < 10^{-
11}$ y$^{-1}$ \cite{dam96}. This value exceeds the limits obtained 
earlier from an analysis of nucleosynthesis processes ($\mid\Delta 
G_F\mid/G_F < 0.06$ ) \cite{reev94} and analysis of the beta decay 
of $^{40}$K ($\mid\dot G_F/G_F\mid < 10^{-10}$ y$^{-1}$) 
\cite{dyson72}. However, it should be kept in mind that these 
limits were obtained under the assumption that all the other 
constants are constant, which makes estimates of this kind less 
reliable. It has not been ruled out that variations of the 
constants are interrelated and the effect due to a change in the 
constant can be compensated by a change in another constant.

\section{$\beta\beta$-decay and time variation of $G_F$}

     Double beta decay is of interest in itself in the problem of 
the change in the fundamental constants with time. The probability 
of ordinary beta decay is proportional to $G_F^2$, while the 
probability of double beta decay goes as $\sim G_F^4$ (since 
$\beta\beta$-decay is of second order in the weak interaction); 
$G_F$ is the Fermi constant. For this reason, if, for example, in 
ordinary $\beta$-decay the effect due to a change in $G_F$ in time 
is compensated by a change in other fundamental constants, then 
this effect could still come through in $\beta\beta$-decay. 
Therefore the study of the time dependence of the rate of 
$\beta\beta$-decay can give additional (and possibly unique!) 
information about the possible change in $G_F$ with time. We 
recall in this connection that the age of minerals and meteorites 
is determined by radioisotopic methods ($\beta$- and $\alpha$-
decay). 
For this reason, when attempts are made to observe a time 
dependence of the rate of $\beta$-decay of $^{40}$K, for example, 
then the change in $G_F$ can be masked by incorrect dating of 
the sample under study.

\section{Comparison of "present" and "past" rate of $\beta\beta$-
decay  for $^{82}$Se and $^{130}$Te}

     Let us compare the rate of $\beta\beta$-decay obtained in 
modern counter experiments to the rate of the same process 
obtained in geochemical experiments, which carry information about 
the rate of $\beta\beta$-decay in the past. Geochemical 
experiments are based on the separation of the products of 
$\beta\beta$-decay from ancient minerals followed by isotopic 
analysis of the products. The observation of an excess quantity of 
daughter isotope attests to the presence of $\beta\beta$-decay of 
the initial isotope and makes it possible to determine its half-
life. 
Minerals containing tellurium and selenium have 
been investigated and the half-lives of $^{130}$Te, $^{128}$Te and 
$^{82}$Se have been measured. Since the age of the 
minerals investigated ranged from $\sim 28$ million years up to 
4.5 
billion years, it is possible in principle to extract from 
geochemical experiments information about the values of $G_F$ in 
the past -- right back to the time when the solar system formed ( 
4.5 billion years ago ). If the value of $G_F$ oscillates with 
time, then these oscillations can be observed.

     Let us examine systematically all the existing experimental 
data.

\underline{1. $^{82}$Se.} The most accurate present-day value of 
the half-life of $^{82}$Se with respect to the $\beta\beta (2\nu)$ 
channel was obtained with the NEMO-2 track detector \cite{arno98}: 
$T_{1/2} = [0.83 \pm 0.10(stat)\pm 0.07(syst)]\cdot 10^{20}$ y. 
The following most precise value was obtained in geochemical 
experiments: $T_{1/2} = (1.30\pm 0.05)\cdot 10^{20}$ y 
\cite{kirs87} ( the average value for 17 independent measurements; 
the age of the samples ranged from 80 million years up to 4.5 
billion years). 
Comparing these results shows that the present-day value of the 
half-life $^{82}$Se is different from the half-life in the past 
(this effect is at the level $\ge 3\sigma$). If this is due to a 
change in the value of the weak-interaction constant, then 
$\Delta G_F/G_F\approx -0.1$, and with the errors taken into 
account the 
possible range of values is approximately $-(0.02-
0.2)$ We note, however, that 
in 
the case of oscillations the interpretation of the experimental 
data becomes much more complicated and depends on the value of the 
period of the oscillations. 

     Let us now to analyse all published results by this time 
(including results
obtained after 1986) and results presented in \cite{kirs87}, mainly, results from 
\cite {kirs67,kirs69,srin73,kirs87,lin86,lin88} 
The results were analysed in three time intervals, $t < 0.1\cdot 10^9$ y, 
$0.17\cdot 10^9 < t < 0.33\cdot 10^9$ y and 
$1\cdot 10^9 < t < 2\cdot 10^9$ y. The following values were obtained:
$T_{1/2} = (0.8\pm 0.15)\cdot 10^{20}$ y, $T_{1/2} = (1.32\pm 0.06)\cdot 10^{20}$ y and $T_{1/2} = (1.28\pm 0.07)\cdot 10^{20}$ y, respectively.
One can see dependence the half-life value with age of minerals.
It means that probability of double beta decay rate of $^{82}$Se now is  
 $\sim 50-70$\% higher than in the past.

 \underline{3. $^{130}$Te, $^{128}$Te.} Only data from geochemical 
measurements are available for these isotopes. Although the ratio 
of the half-lives of these isotopes has been determined to a high 
degree of accuracy ($\sim 3$\%) \cite{bern93}, the absolute values 
of $T_{1/2}$ differ substantially in different experiments. One 
group of authors \cite{lin88,taka66,manu87,taka96} presents the 
values {$T_{1/2}\approx 0.8\cdot 10^{21}$ y for $^{130}$Te and 
$T_{1/2}\approx 2\cdot 10^{24}$ y for  $^{128}$Te, while another 
group \cite{kirs87,bern93} gives $\sim (2.55-2.7)\cdot 10^{21}$ 
y 
and $(7.7\pm 0.4)\cdot 10^{24}$ y, respectively. On closer 
examination one can conclude that, as a rule, experiments with 
"young" minerals ($< 100$ million years) give $\sim (0.7-0.9)\cdot 
10^{21}$ y for $^{130}$Te, whereas experiments on 
"old" ($\ge 1$ billion years) minerals give $\sim (2.5-2.7)\cdot 
10^{21}$ y.

 Let us now again to analyse all published results for $^{130}$Te,
 mainly, results from \cite{ingh50,taka66,kirs67,kirs67a,kirs68,
alex69,srin72a,srin72b,henn75,kirs83,rich86,lin86,lin88,
bern93,taka96}}\footnote{
Uncorrected value from \cite{rich86} was used }.
The results were analysed in two time intervals, $t < 0.1\cdot 10^9$ y 
and 
$1\cdot 10^9 < t < 2.75\cdot 10^9$ y. The following average values
were obtained:
$T_{1/2} = (0.81\pm 0.05)\cdot 10^{21}$ y and 
$T_{1/2} = (1.71\pm 0.04)\cdot 10^{21}$ y, respectively.
One can see again the dependence of half-life value on age of minerals.

In this connection it 
is very important to perform precise measurements of the present-
day 
value of the half-life of $^{130}$Te and $^{82}$Se. Such measurements will 
be performed in the near future in an experiment with the NEMO-3 
track detector \cite{bara97}. It is also obvious that new 
geochemical measurements with samples of different age and 
accuracy $\sim 10$\% are required. Modern mass spectrometry 
makes it possible to perform such measurements with an accuracy of 
several percent (see, for example, \cite{bern93}). The age of the 
samples is also determined, as a rule, with an accuracy of several 
percent. The main uncertainty in geochemical experiments with 
$^{82}$Se and $^{130}$Te is due to the determination of the effective "retention" 
age of daughters $^{82}$Kr and $^{130}$Xe in minerals. To solve this problem it is 
necessary to pick samples which have a well-known geological 
history and for which the retention age of $^{82}$Kr and $^{130}$Xe 
can be accurately determined.

In summary, analysis has shown the following:

     1. A discrepancy exists between the values of the half-life 
of $^{82}$Se which were obtained in modern counter experiments and 
in geochemical measurements.
 
    2. Geochemical measurements on young minerals give lower 
values of $T_{1/2} $($^{82}$Se) as compared to measurements on 
old minerals.

     3. Geochemical measurements on young minerals give lower 
values of $T_{1/2} $($^{130}$Te) as compared to measurements on 
old minerals. That it the same tendency as  for $^{82}$Se.

     These discrepancies can all be explained (at least partially) 
by a change in $G_F$ with time. If this is indeed the case, then 
this will have the most serious consequences for modern physics 
and astrophysics. But, this is why it is necessary to confirm (or 
refute) reliably the reality of these discrepancies. This can be 
done only by performing new and more accurate measurements. We 
propose the following:

 - precise laboratory measurements of the present-day values of 
the $\beta\beta2\nu$-decay half-lives of $^{82}$Se, $^{96}$Zr and 
$^{130}$Te should be performed;

 - new, precise measurements of the half-lives of $^{82}$Se, 
$^{96}$Zr and $^{130}$Te in geochemical experiments should be 
performed; for each isotope it is desirable to perform 
measurements with minerals of different age in order to follow the 
character of the dependence of $G_F$ on the time;

 - the possibility of performing geochemical experiments with 
$^{100}$Mo, $^{116}$Cd, $^{124}$Sn, $^{110}$Pd, $^{150}$Nd and 
$^{76}$Ge should be investigated, and if possible such 
measurements should be performed; this will make it possible to 
enlarge the range of  isotopes investigated, since the half-lives 
of $^{100}$Mo, $^{116}$Cd, $^{150}$Nd and $^{76}$Ge have already 
been measured in direct (counter) experiments \cite {dass95,
arno96,gunt97,silva97}, while the 
half-lives of $^{124}$Sn and $^{110}$Pd can be measured in the 
near future.
 
The best candidate is $^{100}$Mo because of the following 
reasons: 1) maximal $\beta\beta$-decay rate; 2) high concentration 
in natural Mo (9.6\%) and 3) $^{100}$Ru (not gas!) as final 
nucleus. 

\section{Concluding remarks}

We demonstrated that there are discrepancies between results of 
direct and geochemical $\beta\beta$-decay experiments in $^{82}$Se 
and between results for $^{82}$Se and $^{130}$Te with "young" and 
"old" minerals of Se and Te. One of the possible explanation of these 
discrepancies could be the time variation of $G_F$. To check this 
hypothesis new direct and geochemical experiments 
are proposed.
 
In 
fact, $G_F$ is not a "real" fundamental constant. Following, for 
example, ref. \cite{okun90} one can obtain 
that $\eta \sim 1/\sqrt{G_F}$ (where 
$\eta$ is the vacuum expectation value of the Higgs field). It means  that  if $G_F$ is 
increasing with time then $\eta$ is decreasing. Therefor mass of fermions will decrease with time too.

\section{Acknowledgements}
 In conclusion, I wish to express my appreciation to L.B. Okun for 
 a number of helpful remarks.
     
This work was supported by INTAS (grant No. 00-00362).

\end{article}
\end{document}